\documentclass{article}    
\usepackage{epsfig,rotate,spie}

\title{16 x 25 Ge:Ga Detector Arrays for FIFI~LS}

\author{D.~Rosenthal\supit{a}, J.~W.~Beeman\supit{b}, N.~Geis\supit{a}, 
L.~Looney\supit{a}, 
\\
A.~Poglitsch\supit{a}, W.~K.~Park\supit{a}, 
W.~Raab\supit{a}, A.~Urban\supit{a}
\skiplinehalf
\supit{a}Max-Planck-Institut f\"{u}r Extraterrestrische Physik, D-85740 Garching,
Germany
\\
\supit{b}Lawrence Berkeley National Laboratory, Berkeley, CA 94720
}
\authorinfo{Correspondence: rosenthal@mpe.mpg.de; phone: +49 - 89 - 3299 - 
3362; FAX: +49 - 89 - 3299 - 3292}

\begin{document} 

\maketitle 


\begin{abstract}
We are developing two-dimensional $16 \times 25$ pixel detector arrays of both 
unstressed and stressed Ge:Ga photoconductive detectors for far-infrared 
astronomy from SOFIA.
The arrays, based on earlier $5 \times 5$ detector arrays used on the KAO, will
be for our new instrument, the Far Infrared Field Imaging Line Spectrometer 
(FIFI LS).
The unstressed Ge:Ga detector array will cover the wavelength range from 40 to 
120~$\mu$m, and the stressed Ge:Ga detector array from 120 to 210~$\mu$m. The 
detector arrays will be operated with multiplexed integrating amplifiers with 
cryogenic readout electronics located close to the detector arrays. 
The design of the stressed detector array and results of current measurements 
on several prototype 16 pixel linear arrays will be reported. They demonstrate
the feasibility of the current concept.
{\bf This paper does not include Figures due to astro-ph size limitations.
Please download entire file at http://fifi-ls.mpe-garching.mpg.de/fifils.ps.gz.}
\end{abstract}



\section{INTRODUCTION}
For the Far Infrared Field Imaging Line Spectrometer 
(FIFI~LS)\cite{gei98,raa99,loo00}, we need two-dimensional $16 \times 25$ pixel 
detector arrays which cover the wavelength range from 40 to 210~$\mu$m. 
Gallium-doped germanium detectors are proven to be very sensitive in the 
wavelength range of 40 to 120~$\mu$m. Application of 
$\approx$~600~Nmm$^{-2}$ of stress along the [100] crystallographic axis 
shifts the long wavelength cutoff from 120~$\mu$m to approximately
220~$\mu$m\cite{kaz77}. 

Thus, we will use two Ge:Ga detector arrays, one stressed and one unstressed, 
to cover the desired wavelength range. 
The concentrations of 
Gallium dopants will be $1 \times 10^{14}$~cm$^{-3}$ and $2 \times 
10^{14}$~cm$^{-3}$ for the stressed and unstressed arrays, respectively.
The expected dark detector NEP is $\leq 5 \times 10^{-18}$~WHz$^{-1/2}$, which
has been reached with a similar design in a balloon-borne
experiment\cite{hir89}. Some of the operational parameters of the detector
arrays are listed in Tab.~\ref{tab:para}.

\section{ARRAY DESIGN}
The design for the stressed and unstressed detector arrays will be almost
identical. For stability reasons, the {\em unstressed} detector array will 
actually be stressed to about 10~\% of the long-wavelength stress level. 
As a tradeoff between sensitivity and susceptibility to cosmic rays, the size 
of the detector pixels was chosen to be roughly 1~mm$^2$ cross section with an
interelectrode distance of 1.5~mm.
Due to the high reflectivity and low absorption coefficient of Ge:Ga, the 
quantum efficiency of a free standing photoconductor would be very low.
The probabilities for single pass absorption of a FIR photon by detectors this
size are estimated to be 10~\% and 20~\% for unstressed and stressed Ge:Ga,
respectively\cite{wan86}. Therefore, the detectors are located in gold-plated 
integrating cavities with area-filling light cones to maximize the
quantum efficiency. Electrically, the detector housing is maintained at a
constant bias voltage, while the signal ends of the individual pixels are
insulated from the housing by a thin shim of sapphire. 

\subsection{Finite Element Analysis} \label{sec:fem}
\begin{figure}[htb] 
\begin{center}
\caption[]{\em Results of finite element analysis of one detector pixel between
two cylindrical pistons in the case of a perfectly centred (left) and a  
20~$\mu$m-decentred detector pixel for an external force of 500~N. Displayed 
is the stress distribution (in Nmm$^{-2}$) in the direction of the applied 
force. Also shown
is the effect of a pedestal between the pistons and the detector. At right,
the stress distribution of the detector housing is shown.}
\label{fig:fem}
\end{center}
\end{figure}
In order to optimize the stressing mechanism, we studied the stress distribution
of a single detector pixel between two cylindrical steel pistons which apply an
external force of 500~N, as shown in Fig.~\ref{fig:fem}. 
Even in the case of a perfectly centred detector, the 
distribution of stress values within the detector is very inhomogeneous, with 
stress values varying between 405 and 765~Nmm$^{-2}$. This leads to a 
broadening of the spectral response curve and an enhanced probability of 
detector breakage. The stress uniformity within one pixel can be significantly
improved by using a piston of a material with a higher Young's modulus or 
pedestals between the pistons and the detector. In our analysis, the use of 
silver pedestals reduces the variation of stress to a range of 486 to 
507~Nmm$^{-2}$.
As we see from Fig.~\ref{fig:fem} (middle), the range of stress values is 
drastically increased if the detector is slightly (20~$\mu$m) decentred, which
demonstrates the importance of precisely centred detectors. Although silver
pedestals have been used, the stress values vary between 301 and 691~Nmm$^{-2}$.

On the right panel of Fig.~\ref{fig:fem}, the stress distribution for
the detector housing is shown.
The highest stress levels occur at the ``c-shaped'' clamp, while the material 
near the detector stack remains unaffected. 
  
\subsection{Stressing Mechanism}
Each detector is placed within about 20~$\mu$m of the centre of its cavity. 
As the finite element analysis (FEA) of the previous section has shown, this 
positional accuracy is required to avoid inhomogeneous stress, which would leadin
to inhomogeneous responsivity and increased probability of pixel breakage. 
The edges of the detectors face
\begin{figure}[htb]
\begin{center}
\caption[]{\em Drawing of the detector housing of one 1 x 16 pixel linear
stressed Ge:Ga array. The light cones are tilted in the focal plane to match
the angle of incoming light from the pupil plane at a distance of 240~mm. The
enlarged section shows a single detector cavity consisting of the detector,
the combination of stress pistons, the CuBe pad, and the sapphire pad.
}
\label{skizze}
\end{center}
\end{figure}
\begin{table} 
\caption{
Parameters of the Detector Arrays}
\begin{center}
\begin{tabular}{l|cc}
\hline
\hline
\multicolumn{1}{c}{detector} &
\multicolumn{1}{c}{unstressed} &
\multicolumn{1}{c}{stressed} \\
\multicolumn{1}{c}{type} &
\multicolumn{1}{c}{Ge:Ga} &
\multicolumn{1}{c}{Ge:Ga} \\
\hline
\rule[-1mm]{0mm}{4mm}wavelength range& 40 -- 120~$\mu$m & 120 -- 210~$\mu$m \\
\rule[-1mm]{0mm}{4mm}size  & $16 \times 25$& $16 \times 25$ \\
\rule[-1mm]{0mm}{4mm}bias field& 10 -- 20~Vm$^{-1}$ & 100 -- 200~Vm$^{-1}$\\
\rule[-1mm]{0mm}{4mm}operating temperature & 1.8~K& 3.5~K \\
\rule[-1mm]{0mm}{4mm}doping concentration& $1 \times 10^{14}{\rm cm^{-3}}$&$2 
\times 10^{14}{\rm cm^{-3}}$ \\
\hline
\end{tabular}
\end{center}
\label{tab:para}
\end{table} 
the entrance holes of the cavities, to ensure that the first pass
of reflected radiation is trapped by the integrating cavity and not reflected
directly back out through the entrance aperture. 
The entire detector array will consist of 25 linear $1 \times 16$ pixel 
detector arrays, each equipped with its own linear light cone array. 
In Fig.~\ref{skizze}, one linear stressed Ge:Ga detector array
is sketched. The detector housing as well as the light cone arrays are 
machined by sparc erosion, which ensures high precision without introducing
mechanical stress. As for the previous $5 \times 5$ FIFI detector 
array\cite{pog91,sta92}, one screw at the top of the array serves as the 
stressing mechanism. To gradually increase the stress to the detectors when 
the stressing screw is turned, the horizontal slit is designed to create a 
spring mechanism.
The spring constant at room temperature for the stressed detector array is
$\approx~513$~Nmm$^{-1}$ and will be about $\approx~50$~Nmm$^{-1}$ for the 
unstressed array. 

The detector housing is constructed of a high strength aluminum alloy 
(7075 Aluminum T6) which ensures a high mechanical stability and a good thermal
conductivity. Due to a differential thermal contraction between the housing and
the components of the detector stack, the stress to the detectors is increased 
during cool-down. For the stressed array, this increase is estimated to be
about 100~N. 

The vertical slit decouples the detector stack from any distortions caused by 
the stress application. 
The ball-and-socket pivot design of the tungsten carbide pistons compensates for
non-parallel surfaces. To provide a
uniform stress along the stack of detectors, the upper and lower
parts of this combination are machined from spheres with diameter matching 
those of the detector cavities of 3~mm, so that they can rotate slightly 
without conducting forces into the housing. 
Due to the high Young's modulus of tungsten carbide, the bending of the pistons
around the corners of the detectors is minimized.
The signal end of a detector is in contact with the pedestal of a CuBe pad.
Both the high Young's modulus of the tungsten carbide pistons and the pedestals
of the CuBe pads lead to an improved stress homogenity within the pixels (see
section~\ref{sec:fem}). The CuBe pad is electrically insulated from the housing
by a thin sheet of sapphire. The signal wire is soldered to the CuBe pad and 
led to the back end of the detector housing, where it is soldered to a 
connector. The connector is linked to the cryogenic readout electronics (CRE) 
located at the back of the detector housing.

\subsection{Light Cones} \label{sec:cone}
\begin{figure}[htb]
\begin{center}
\caption[]{\em left: Photograph of a prototype of the two types of light cones 
which are under study. right: Raytracing result to determine the transmission 
of the two types of light cones using the program ``OptiCAD''. The dotted line 
indicates the extent of the pupil. 
The transmission was determined by placing a point light source at a 
certain angle and counting the percentage of transmitted rays from a ray bundle
which fully illuminated the entrance aperture of each light cone.}
\label{fig:cone}
\end{center}
\end{figure}
To collect all of the light in the focal plane and to feed it into the
appropriate detector cavity, funnel-shaped light cones are used. The back end of
each linear light cone array forms a part of the integrating cavity (see  
Fig.~\ref{skizze}). Like the cavities, the light cones are gold plated. 
The light cones are individually tilted in the focal plane to match the angle of
incoming light from the pupil plane. Due to an easier and therefore cheaper
production, we will use straight light cones which have been successfully used 
on the KAO with FIFI and not the parabolical Winston-type cones. 
In order to optimize the light cone arrays, we studied two 
different types of light cones (Fig.~\ref{fig:cone}) whose parameters are 
listed in Tab.~\ref{tab:cone}. 
Since the coupling holes are the main loss source of the 
detector cavities, we attempt to reduce their size as much as possible to 
optimize the quantum efficiency. 
On the other hand, the diameter of the coupling hole should be of order at least
a few $\lambda$ to minimize diffraction losses. To what extent type~II with a
coupling hole diameter of 0.5~mm is already affected by diffraction losses is
not yet clear and has to be determined. On the right panel of 
Fig.~\ref{fig:cone}, the simulated transmission for both types is plotted. Ray 
bundles from solid angles exceeding those of the pupil are no longer entirely 
transmitted, allowing the light cones to help reduce straylight. In this 
respect, type~II
is somewhat more efficient: it totally rejects light from light sources 
at angles $\geq 10^\circ = \alpha_{\rm cr}$ from the optical axis
of the light cone, whereas $\alpha_{\rm cr} = 15$ for type~I. However, this 
raytrace simulation uses geometrical optics and does not account for 
diffraction losses.

\begin{table} 
\caption{
Parameters of the two types of light cones}
\begin{center}
\begin{tabular}{l|cc}
\hline
\hline
\multicolumn{1}{c}{\rule[-2mm]{0mm}{6mm}} & 
\multicolumn{1}{c}{type I} &
\multicolumn{1}{c}{type II} \\
\hline
diameter of coupling hole  & 0.72~mm & 0.5~mm \\
effective entrance hole$^a$ & $3.6 \times 3.6$~mm$^2$ & $3.4 \times 
3.4$~mm$^2$ \\
length & 19.9~mm & 30.9~mm \\
distance to pupil$^b$ & 240.1~mm & 229.1~mm \\ 
\hline
\end{tabular}
\parbox{9cm}{$^a$ The circular 
entrance apertures of 5.1 and 4.9~mm diameter are cut to these
square entrance apertures so as to completely cover the focal plane.\\
$^b$ Distance of aperture from the (previous) pupil.}  
\end{center}
\label{tab:cone}
\end{table} 

\subsection{Readout Electronics} \label{sec:readout}
\begin{figure}[htb]
\begin{center}
\caption[]{\em Schematic of the cryogenic readout electronics (CRE).}
\label{readout}
\end{center}
\end{figure}
The purpose of the cryogenic readout electronics (CRE), together with some 
passive components, is to amplify and multiplex the signal of the 16 detectors
of one linear array.
The CRE is a specially designed CMOS circuit under development for 
FIRST-PACS\cite{die96,pog99}, operating at liquid helium temperature or lower
(see Fig.~\ref{readout}).
The detector current of each pixel is read out by an integrating amplifier. A
sample-and-hold stage acts as analog memory between the integrator and the 
multiplexer circuit. All channels of one complete detector column are sampled
at the same time and switched sequentially to the output of the CRE by the 
sample-and-hold circuit. 

The complete read out of each $16 \times 25$ pixel
photoconductor array is done by a 16-bit BiCMOS A/D converter stage. Very 
careful shielding, including output signal feedback on the shield, is used to
minimize the load on the cryogenic output stage and the crosstalk between the
analog signal lines to the converter stage.
A digital multiplexer circuit provides parallel/serial conversion of the 16-bit
data words generated by the A/D converter stage for both photoconductor arrays.
The resulting 8~Mbit per second serial data stream (at 10~kHz maximum sampling
rate) is transmitted to the data acquisition system via fiber obtics.

\section{PERFORMANCE MEASUREMENTS}
Several prototype linear $1 \times 16$ pixel arrays based on the above design
have been succeessfully assembled and tested. A photograph of one of the 
assembled arrays is shown in Figure~\ref{gehaeuse}.
\begin{figure}[htb]
\begin{center}
\caption[]{\em Photograph of one prototype $1 \times 16$ linear detector
array. On the left is the whole assembly, including the light cones, and on the
right is a detailed view of one detector cavity.}
\label{gehaeuse}
\end{center}
\end{figure}
 
\subsection{Stress Uniformity}
\begin{figure}[htb]
\begin{center}
\caption[]{\em left: Room temperature resistance as a function of the applied
stress measured for five detectors of one linear array. The indices indicate the
positions (counted from the stressing screw) of the detectors within 
the stack. right: Relative spectral response of four detectors measured at 2~K.}
\label{uniform}
\end{center}
\end{figure}
To verify the functionality of the described stressing mechanism, we measured
the room temperature resistance as a function of the applied stress, as shown in
the left part of Fig.~\ref{uniform}. The decrease of the resistance with
increasing stress is very homogeneous along the stack of detectors, and no 
stress gradient is noticeable.

The relative spectral responses of four detectors at different positions within 
the stack measured with a Fourier-Transform Spectrometer (FTS) for a stress of
540~Nmm$^{-2}$ also match quit well (Fig.~\ref{uniform}). The 
cut-off wavelengths of the four detectors agree especially well.

\subsection{Responsivity and Noise Equivalent Power}

\begin{figure}[htb]
\begin{center}
\caption[]{\em Measured responsivity (left) and noise equivalent power 
(NEP, right) vs. the applied bias field for three detectors of one linear 
array compared with the 
performance of the FIFI detector array\cite{sta92}. Using a 20~K blackbody as
the light source, the Fabry-Perot spectrometer was tuned to a centre wavelength 
of 170~$\mu$m with a spectral resolution $\lambda /\Delta \lambda$ of $\approx
50$, which produced a photon power of $3 \times 10^{-13}$~W per pixel. 
Unfortunately, in our preliminary measurements the detectors received a 
background four times higher due to stray light. 
Under the assumption of background-limited 
performance, the measured NEP was therefore extrapolated to the lower
background expected for FIFI~LS ($\lambda = 
170~\mu$m, $\lambda /\Delta \lambda = 2000$) on SOFIA.      
}
\label{resnep}
\end{center}
\end{figure}

A setup of two blackbodies at 4~K and 20~K, one linear stressed detector array,
and a Fabry-Perot interferometer, tuned to a centre wavelength of 
170~$\mu$m at a resolving power of about 50, was used to measure the 
responsivity and NEP of the detectors\cite{urb99}. The parameters were set to
produce a photon power of $3 \times 10^{-13}$~W per pixel which corresponds
to the expected background with FIFI~LS ($\lambda = 170~\mu$m, $\lambda /
\Delta \lambda = 2000$) on SOFIA. Unfortunately, in our
preliminary measurement the detectors received a photon background four times
higher due to unwanted stray light. Since we also lacked the 
readout electronics described in section~\ref{sec:readout}, we used the 
transimpedance amplifiers (TIAs) with GaAs - FET's used for the FIFI array. 
With that, only a few detectors could be tested.  

In Fig.~\ref{resnep} we compare the responsivity and NEP measured for our
detectors and for the FIFI array, measured with narrow-band filters
centred at 163~$\mu$m at a photon background of $2.39 \times 10^{-13}$~W per
pixel\cite{sta92}. 
As shown in Fig.~\ref{resnep}, the responsivity of our detectors at a given bias
field is lower than for the FIFI array. However, as we see from the NEP
measurement, this effect is effect is compensated by our ability to apply a 
higher bias field. 
The NEP of the detectors is almost constant over the considered 
range of bias fields, whereas the NEP of the FIFI detectors rises steeply 
for bias fields $\geq 13$~Vm$^{-1}$ where impact ionization leads to increased
noise.

Even with a background four times higher, above a bias field of about
7~Vm$^{-1}$, our measured NEP lies below that measured for the FIFI array. Under
the assumption of a background-limited performance of the detectors and no noise
contribution from the 
readout electronics, we extrapolated the measured NEP to the desired photon 
power of $3 \times 10^{-13}$~W per pixel. The extrapolated NEP is well below 
that measured for the FIFI array, which may be due to the improved cavity 
design and the resulting enhanced quantum efficiency. 
The $NEP_{\rm BLIP}$ for background limited performance can be expressed as
\begin{equation} \label{eq:nep}
  NEP_{\rm BLIP} = \frac{h\nu}{\eta} \; \left( \frac{2^3 A \Omega}{\lambda^2} 
  \Delta \nu t \epsilon \eta \; \frac{1}{1 - {\rm e}^{h \nu /kT}} \left(
  1 + t \epsilon \eta \, \frac{1}{1 - {\rm e}^{h \nu /kT}} \right)  
  \right)^{1/2},
\end{equation}
where t and $\epsilon$ are the transmission and emissivity of a blackbody at a 
temperature T, and $\eta$ is the quantum efficiency of the detectors.
With the measured NEP we used Eq.~\ref{eq:nep} to calculate a quantum 
efficiency $\eta = 45~\%$, which is a substantial improvement over the average 
of 19~\% reported for the FIFI detector array\cite{sta92}.

The results of our first measurements are encouraging. However, we need more
measurements on a larger number of detectors to confirm these first results.

\section*{ACKNOWLEDGMENTS}
We are thankful to H.~Dohnahlek and G.~Kettenring for the design work, and to
A.~J.~Baker for careful reading of the manuscript.


\begin{thebibliography}{99}

    \bibitem{gei98}
    Geis, N., Poglitsch, A., Raab, W., Rosenthal, D., Kettenring, G., Henning,
    T., \& Beeman, J. in {\em Infrared Astronomical
    Instrumentation}, SPIE proc., Fowler, A.~M. (eds.) 1998, p.~973

    \bibitem{raa99}
    Raab, W., Geis, N., Looney, L., Poglitsch, A., Rosenthal, D., Urban,
    A., Henning, T. \& Beeman, J.~W. in {\em The International Symposium on
    Optical Science, Engeneering and Instrumentation}, SPIE proc., Fowler, A.~M.
    (eds.) 1999, p.~973

    \bibitem{loo00}
    Looney, L.~W., Geis, N., Genzel, R., Park, W.~K., Poglitsch, A.,
    Raab, W., Rosenthal, D., \& Urban, A. 2000, this volume 

    \bibitem{kaz77}
    Kazanskii, A., Richards, P.~L., \& Haller, E.~E. 1977, Ap. Phys. Lett.,
    31, 496

    \bibitem{hir89}
    Hiromoto, N., Itabe, T., Argua, T., Okuda, H., Matsuhara, H., Shibai, H.,
    Nakagawa, T., \& Saito, T. 1989, IR Phys., 29, 255

    \bibitem{wan86}
    Wang, J.-Q., Richards, P.~L., Beeman, J.~W., Haegel, N.~M., \&
    Haller, E.~E. 1986, Appl. Optics, 25, 4127

    \bibitem{pog91}
    Poglitsch, A., Beeman, J.~W., Geis, N., Genzel, R., Haggerty, M.,
    Haller, E.~E., Jackson, J., Rumitz, M., Stacey, G.~J., \& Townes, C.~H.
    1991, Int. J. Inf. Mill. Wav., 12, 859

    \bibitem{sta92}
    Stacey, G.~J., Beeman, J.~W., Haller, E.~E., Geis, N., Poglitsch, A.,
    \& Rumitz, M. 1992, Int. J. Inf. Mill. Wav., 13, 1689

    \bibitem{die96}
    Dierickx, B., Seijnaeve, J., \& Schefferin, D. in {\em Submillimetre and
    Far-Infrared Space Instrumentation}, Rolfe, E.~J. (ed.), Proc. ESA 
    Symposium ESA SP-388,1996, p.~61

    \bibitem{pog99}
    Poglitsch, A., Waelkens, C., \& Geis, N.in {\em The International Symposium
    on Optical Science, Engeneering and Instrumentation}, SPIE proc., 
    Fowler, A.~M.(eds.) 1999, p.~221

    \bibitem{urb99}
    Urban, A., 1999, diploma thesis, Ludwig-Maximilians-Universit\"at M\"unchen

\end{thebibliography}
\end{document}